\documentclass[aps,prb,twocolumn,superscriptaddress]{revtex4}
\usepackage{amsmath,amssymb}
\usepackage{graphicx}
\usepackage{epstopdf}
\usepackage{bm}

\begin{document}


\preprint{LA-UR-09-06981}

\title{$^{235}$U nuclear relaxation rates in an itinerant antiferromagnet USb$_2$}


\author{S.-H. Baek}
\affiliation{Los Alamos National Laboratory, Los Alamos, New Mexico 87545, USA}
\author{N. J. Curro}
\affiliation{Department of Physics, University of California, Davis, California 95616, USA}
\author{H. Sakai}
\affiliation{Los Alamos National Laboratory, Los Alamos, New Mexico 87545, USA}
\affiliation{Advanced Science Research Center, Japan Atomic Energy Agency,
Tokai, Ibaraki 319-1195, Japan}
\author{E. D. Bauer}
\affiliation{Los Alamos National Laboratory, Los Alamos, New Mexico 87545, USA}
\author{J. C. Cooley}
\affiliation{Los Alamos National Laboratory, Los Alamos, New Mexico 87545, USA}
\author{J. L. Smith}
\affiliation{Los Alamos National Laboratory, Los Alamos, New Mexico 87545, USA}

\date{\today}

\begin{abstract}
$^{235}$U nuclear spin-lattice ($T_1^{-1}$) and spin-spin ($T_2^{-1}$) 
relaxation rates in the itinerant antiferromagnet USb$_2$ are reported as a 
function of temperature in zero field.
The heating effect from the intense rf pulses that are necessary for the $^{235}$U 
NMR results in unusual complex thermal recovery of the nuclear  
magnetization which does not allow measuring $T_1^{-1}$ directly.  
By implementing an indirect method, however, we successfully extracted 
$T_1^{-1}$ of the $^{235}$U.   
We find that the temperature dependence of $T_1^{-1}$ for both $^{235}$U and 
$^{121}$Sb follows the power law ($\propto T^n$) with the small exponent 
$n=0.3$ suggesting that the same relaxation  
mechanism dominates the on-site and the ligand nuclei, but an anomaly at 5 K 
was observed, possibly due to the change in the transferred hyperfine 
coupling on the Sb site.  
\end{abstract}

\maketitle

\section{Introduction}

In actinide-based materials, $5f$-electrons often exhibit itinerant and 
localized behavior simultaneously, which is in contrast to the 
usually localized $4f$ electrons in the rare earth compounds.  The unique nature 
of the $5f$ electrons has been known to be the origin of various unusual physical properties
found in actinide based materials such as unconventional 
superconductivity, non-Fermi liquid behavior, and multipolar ordering.  
However, since the degree of the $5f$ localization is highly sensitive to the specific 
ligand atoms and the crystal structure, the nature of $5f$ electrons is 
not easily elucidated even in a single compound.

In principle, nuclear magnetic resonance (NMR) is an ideal method to 
investigate $5f$ electrons by  
probing on-site actinide nuclei ($^{235}$U, $^{237}$Np, $^{239}$Pu) since 
they are directly influenced by $5f$ electrons both  
dynamically and statically. However, NMR in the actinide nuclei is extremely difficult.
For $^{235}$U, for instance, the tiny nuclear gyromagnetic ratio
$\gamma_n=0.784$ MHz/T and very low natural abundance (0.72\%) present 
significant challenges for detecting the NMR.  Furthermore, the fast spin fluctuations 
of the $5f$ electrons require an ordered state in which the spin fluctuations are 
sufficiently suppressed to allow the detection of NMR signal.  Despite these 
difficulties, $^{235}$U NMR was successfully carried 
out recently\cite{ikushima98,kato04} in an insulating UO$_2$ and an itinerant USb$_2$ 
in their antiferromagnetically ordered states. While $^{235}$U 
nuclear relaxation rates were measured in detail in UO$_2$, these quantities 
were not measured in USb$_2$ that 
is highly itinerant.  Motivated by the absence of $T_1^{-1}$ in metallic 
U-based materials, we investigated the $^{235}$U nuclear relaxation rates in USb$_2$.   

USb$_2$ is a member of the uranium dipnictides, UX$_2$ family (X = P, As, Sb, Bi), 
which is characterized by strong magnetic and electronic anisotropies, and 
the hybridization of the $5f$ electrons with the conduction 
electrons.\cite{amoretti84,tsutsui04,henkie92,lebegue06} 
USb$_2$ crystallizes in the tetragonal Cu$_2$Sb type structure (space group: $P4/nmm$) 
and undergoes antiferromagnetic transition at $T_N=203$ K with an ordered 
moment of 1.88 $\mu_B$.\cite{leciejewicz67}   The magnetic unit cell is 
doubled along the $c$-axis with respect to the chemical unit cell due
to the sequence of alternating FM layers ($\uparrow\downarrow\downarrow\uparrow$), as  
depicted in Fig.~1. 
dHvA experiment\cite{aoki99} detected the 
two-dimensional Fermi surfaces which are in agreement with the band 
calculations,\cite{lebegue06} and the dual nature of the $5f$ 
electrons was confirmed by ARPES study\cite{guziewicz04} from very narrow strongly 
dispersive bands at the Fermi level with $5f$ character.

In this paper, we report the $^{235}$U
nuclear spin-lattice ($T_1^{-1}$) and spin-spin ($T_2^{-1}$) 
relaxation rates in $^{235}$U enriched USb$_2$. 

\section{Sample preparation and experimental details}

We have grown single crystals of USb$_2$ enriched with $^{235}$U 
(93.5 \% enrichment) using flux growth in excess Sb.  The $^{235}$U was arc-melted 
prior to the flux growth to remove the high vapor pressure daughters, radium 
in particular.  
Most of the USb$_2$ 
produced was in the form of a single crystal weighing approximately 340 mg.  
Because the rf penetration depth is small an increase in signal strength can 
be accomplished by powdering the sample. The crystal was broken into pieces 
and $\sim$ 100 mg of material was ground into powder using an agate mortar and 
pestle.  A pickup coil with an inside diameter of $\sim$2 mm was cast into an epoxy 
block and after curing a cylindrical sample space was drilled into the epoxy 
within the ID of the pickup coil.  A 2 micron pore size stainless steel frit 
was glued over one end of the sample space.  The USb$_2$ powder was funneled into 
the open end of the sample space and a then a second frit glued on to close 
the open end.  The frits allow thermal contact of the powder with the 
cryogenic fluid/gas and prevent the radioactive 
material from spreading into the apparatus. 

$^{235}$U and $^{121}$Sb NMR were performed in zero field in the temperature 
range 1.5--80 K.  
The NMR spectra were obtained by integrating averaged spin echo signals as a function of
frequency, and the spin-lattice relaxation rates ($T_1^{-1}$) were
measured by acquiring Hahn echoes following various delays after a
saturation pulse, i.e., $\pi/2-t-\pi/2-\tau-\pi$, where $t$ represents the variable 
delay, $\pi/2$ pulse was about 10 $\mu$s, and the repetition time 
longer than 1 s was used to reduce the heating effect.   
In order to extract $T_1$, we fit the raw data with the 
appropriate relaxation functions.  The value of the nuclear spin-spin relaxation rate 
($T_2^{-1}$) was obtained by monitoring the spin-echo amplitude, $M(2\tau)$, as 
a function of $2\tau$ between the first pulse and the echo.  The $M(2\tau)$ 
were fitted to an exponential decay curve $M(0)\exp(-2\tau/T_2)$.  

\section{Experimental results and discussion}

\begin{figure}
\centering
\includegraphics[width=0.45\textwidth]{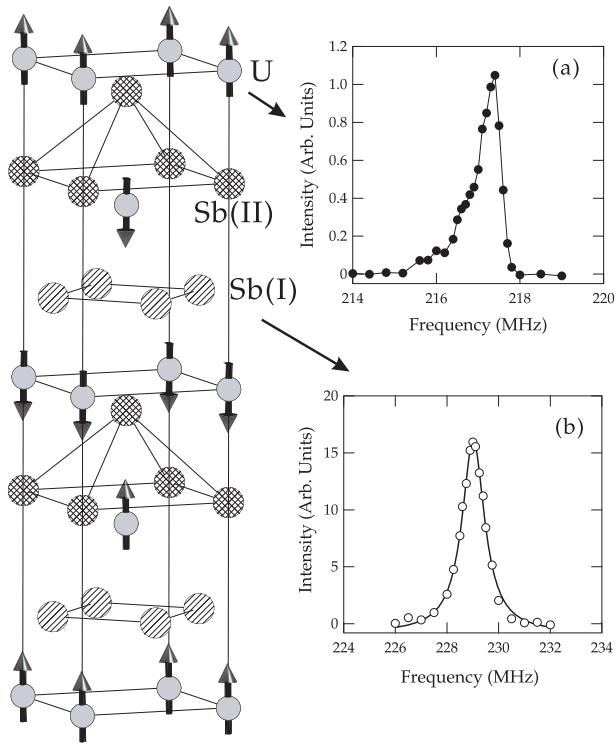}
\caption{The crystallographic and magnetic structure of USb$_2$. Arrows 
indicate the direction of the ordered moments along $c$-axis. Ferromagnetic 
planes are coupled ferromagnetically through Sb(I) plane but 
antiferromagnetically through Sb(II) plane.  (a) $^{235}$U 
NMR spectrum and (b) second satellite spectrum of $^{121}$Sb(I) in zero field.} 
\label{fig:structure}
\end{figure}

From the detailed spectra of $^{121,123}$Sb and $^{235}$U given in 
ref.~\onlinecite{kato04}, we were able to confirm $^{235}$U  
signal at 217.4 MHz and the second satellite transition ($5/2 \leftrightarrow 
3/2$) of $^{121}$Sb(I) at 229 MHz as shown in Fig.~1 (a) and (b). 
While $^{121}$Sb spectrum has a Lorentzian shape, the U spectrum is asymmetric 
with shoulders in low frequency side. Since there is only one U 
crystallographic site in the unit cell, we suggest that a transferred 
hyperfine coupling from nearest neighbor U sites 
may lead to different inequivalent U sites in the complex magnetic
structure. Indeed, the alternating ferromagnetic U planes suggest that different 
transferred hyperfine coupling may result from the different interlayer magnetic 
interactions either through Sb(I) plane or Sb(II) plane. 
We will not present further analysis of this complicated $^{235}$U spectrum 
in this paper and, instead, we focus on the nuclear relaxation rates of 
$^{235}$U, which have never been directly measured in an itinerant magnetic material.

\subsection{Thermal recovery of $^{235}$U nuclear magnetization}

\begin{figure}
\centering
\includegraphics[width=0.45\textwidth]{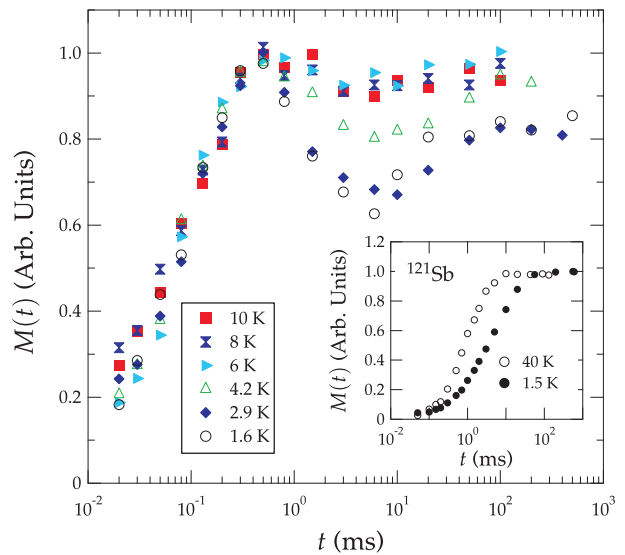}
\caption{Recovery of nuclear magnetization $M(t)$ as a function of time $t$ 
with varying temperature $T$.  $M(t)$ shows an oscillating behavior with 
varying $t$, forming 
a local minimum near $t=5$ ms.  For $t<1$ ms, $M(t)$ is almost independent of 
$T$, while the local minimum becomes deeper with decreasing temperature 
maintaining the same position in time. INSET: For comparison, the recovery 
curves for $^{121}$Sb are shown.  } \label{fig:M}
\end{figure}

As we try to measure $^{235}T_1^{-1}$, it turns out that the recovery of the 
nuclear magnetization, $M(t)$, is very unusual as shown in Fig.~2. 
With our experimental conditions, we were unable to make the full saturation 
of the line and there are  
always sizable signal more than $20$ \% with regard to the full amplitude. 
Also $M(t)$ as a function of temperature $T$ does not 
change below $t<0.5$ ms but it reveals an oscillating behavior with the minimum 
near $t=5$ ms. With decreasing $T$, the minimum 
becomes deeper without the change of its position in $t$.

We speculate that the unusual behavior of $M(t)$ is due to the strong 
rf pulse in the sample coil, which may produce considerable heat causing complex thermal 
recovery of the nuclear magnetization. 
To flip a nuclear spin, we apply a $\pi/2$ pulse that satisfies 
the relation $\gamma_n H_1 \tau = \pi/2$, where $H_1$ corresponds to the rf 
strength (power) and $\tau$ the duration of the pulse.  Since $\gamma_n=0.784$ MHz/T of 
$^{235}$U is one order of magnitude smaller  
than typical nuclei (e.g., 10.189 MHz/T for $^{121}$Sb), the total energy 
transferred to the coil, $H_1 \tau$,  
should be large correspondingly. Moreover, the cooling power is substantially 
reduced in the measurement since the sample is located inside an epoxy block 
to prevent the contamination. 
Thus, the heating effect may be a consequence 
from the experimental limitations with regard to the $^{235}$U nuclei. 

In order to account for the heating effect, we measured two data sets with different 
experimental conditions.  One is measured through the usual sequence for 
measuring $T_1$.  In an other sequence, we apply 
the saturating $\pi/2$ pulse at off-resonance frequency 
$\omega_\text{off}$ as depicted in the inset of Fig.~3 (a).  The detecting 
pulses consisting of $\pi/2-\pi$ are the same in both sequences.  
Here, the $Q$-factor of the tank circuit is ensured to be low enough to cover 
$\omega_\text{off}$ so  
that the rf power applied at $\omega_\text{off}$, $(\pi/2)_\text{off}$, is 
fully transferred to the  
sample coil producing the similar amount of heat as the rf pulse applied at the 
resonance frequency $\omega_\text{on}$.  
If there is no heating effect, the second pulse sequence with 
$(\pi/2)_\text{off}$ should give rise to a constant 
magnetization $M(t)=M_0$, since the $(\pi/2)_\text{off}$ pulse does not flip the 
nuclei. Using this procedure, the thermal recovery of the magnetization 
$M(t)_\text{therm}$ oscillates in a similar fashion with the total recovery of 
the magnetization $M(t)_\text{tot}$ as shown in Fig.~3 (a). 

We treat $M(t)_\text{therm}$ as the fully recovered constant value $M_0$ 
at each time so that the nuclear relaxation function can be written as 
\begin{equation}
\label{}
 R(t) = 1-M(t)_\text{tot}/M(t)_\text{therm}. 
\end{equation}
This accounts for not only the heating effect inside the sample coil but also any 
possible artificial effect originating from the power amplifier or the receiver.  
The corrected relaxation data are shown in Fig.~3 (b), and we fit the data with 
the relaxation function for the central transition of $I=7/2$ (solid lines),
\begin{equation}
\begin{split}
\label{eq:fit}
R(t) =&  
\frac{1}{84}\exp\left(-\frac{t}{T_1}\right)+\frac{3}{44}\exp\left(-\frac{6t}{T_1}\right)\\
&+\frac{75}{364}\exp\left(-\frac{15t}{T_1}\right) 
+\frac{1225}{1716}\exp\left(-\frac{28t}{T_1}\right).
\end{split}
\end{equation}
Here we assume a large quadrupole frequency $\nu_Q$ which is 
estimated to be $\sim 140$ MHz from M\"{o}ssbauer spectroscopy\cite{tsutsui01} 
so that the rf irradiation induces the central transition only.  Also we 
expect that the spectral diffusion, if any,
does not affect the obtained $T_1$ values since it should occur at times that are much shorter 
than $T_1$. 
Note that the correct scaling behavior in $T_1^{-1}$
between $^{121}$Sb and $^{235}$U in Eq.~(\ref{eq:T1scaled}), as discussed 
below, supports the validity of Eq.~(\ref{eq:fit}). 

\begin{figure}
\centering
\includegraphics[width=0.45\textwidth]{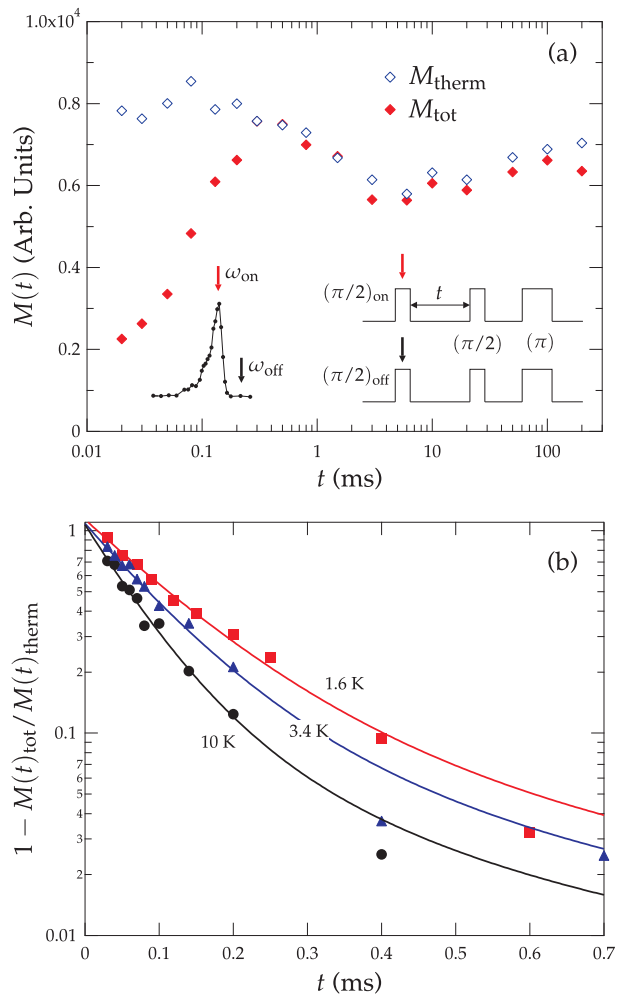}
\caption{(a) Total and thermal magnetization $M(t)_\text{tot}$ and 
$M(t)_\text{therm}$, respectively.  
For $M_\text{tot}$, the typical pulse sequence for measuring $T_1$ 
was used (upper diagram in inset).  For $M_\text{therm}$, a saturating $\pi/2$ pulse was 
applied at an off-resonant frequency $\omega_\text{off}$ in 
order to produce heating effect only, without affecting the NMR line. 
(b) Corrected relaxation function of nuclear magnetization, 
$1-M(t)_\text{tot}/M(t)_\text{therm}$,  which is free of heating effect, for 
three selected temperatures.  Solid lines are fits with Eq.~(\ref{eq:fit}). } 
\label{fig:relaxation}
\end{figure}

\subsection{Nuclear relaxation rates, $T_1^{-1}$ and $T_2^{-1}$}

The $^{235}$U nuclear spin-lattice relaxation rate $^{235}T_1^{-1}$ 
as a function of temperature is shown in  
Fig.~4 (a). 
The data can be fit by a power law $T^{0.3}$ in the measured temperature range.  
$^{121}$Sb nuclear spin-lattice relaxation rate $^{121}T_1^{-1}$ also shows 
the same power law behavior with a factor of 4 smaller prefactor than 
that of $^{235}T_1^{-1}$.  In general, $T_1^{-1}$ due to the spin fluctuations 
is given by\cite{moriya63} 
\begin{equation}
\label{eq:T1}
T_1^{-1} \approx 2T\gamma_n^2 A_\text{hf}^2 \frac{\sum_q \chi_\perp''(q,\omega_0)}{\omega_0},
\end{equation}
where $A_\text{hf}$ is the hyperfine 
coupling constant at $q=0$, $\chi_\perp''$ is the  
imaginary part of the $q$-dependent dynamic susceptibility at the nuclear 
Larmor frequency $\omega_0$ that represents the spin 
fluctuations in the perpendicular plane. Since 
$\sum_q \chi_\perp''(q,\omega_0)$ should be the same for both nuclei, the 
following relation should hold:  
\begin{equation}
\label{eq:T1scaled}
\frac{^{235}T_1^{-1}}{^{121}T_1^{-1}} = 
\frac{^{235}(\gamma_n A_\text{hf})^2}{^{121}(\gamma_n A_\text{hf})^2},
\end{equation}
where $A_\text{hf}$ is 5.69 T/$\mu_B$ for $^{121}$Sb and 147.5 T/$\mu_B$ for 
$^{235}$U.\cite{kato04} Indeed, the  
experimental values $^{235}T_1^{-1}$ and $^{121}T_1^{-1}$ are well scaled 
according to Eq.~(\ref{eq:T1scaled}) above 5 K as shown in the inset of Fig.~4(a). 
The slight difference between the two data may 
be due to systematic error from to the \textit{indirect} way of 
acquiring $^{235}T_1^{-1}$. However, we find that $^{235}T_2^{-1}$ and $^{121}T_2^{-1}$ 
data are also scaled with the same ratio between the two
$T_1^{-1}$ data sets at high temperatures. This suggests that the spin 
fluctuations dominate both $T_1^{-1}$ and $T_2^{-1}$ for the two nuclei. 
Since both $T_1^{-1}$ and $T_2^{-1}$ are scaled with the same ratio, we 
argue that the slight difference of $\sum_q \chi_\perp''(q,\omega_0)$ may be 
attributed to an additional contribution to the relaxation rates other than
the spin fluctuations, probably, due to the lattice vibrations (phonons) which 
are not necessarily the same for both nuclei.

In the case that the spin fluctuations dominate, $T_2^{-1}$ may be 
written as
\begin{equation}
\label{eq:T2}
T_2^{-1} \approx T \gamma_n^2  
A_\text{hf}^2 \frac{\sum_q 
\chi_\parallel''(q,\omega_0)}{\omega_0}+\frac{1}{2}T_1^{-1}+T_{2,\text{dip}}^{-1},
\end{equation}
where $T_{2,\text{dip}}^{-1}$ is the dipolar term which is estimated to be 
$\sim 10^{3}$ s$^{-1}$ for both $^{121}$Sb and $^{235}$U by summing up the dipolar 
contribution from the nearest neighbors.\cite{abragam-nmr} 
In the meantime, we assume that the hyperfine coupling constant $A_\text{hf}$ is isotropic. 

\begin{figure}
\centering
\includegraphics[width=0.45\textwidth]{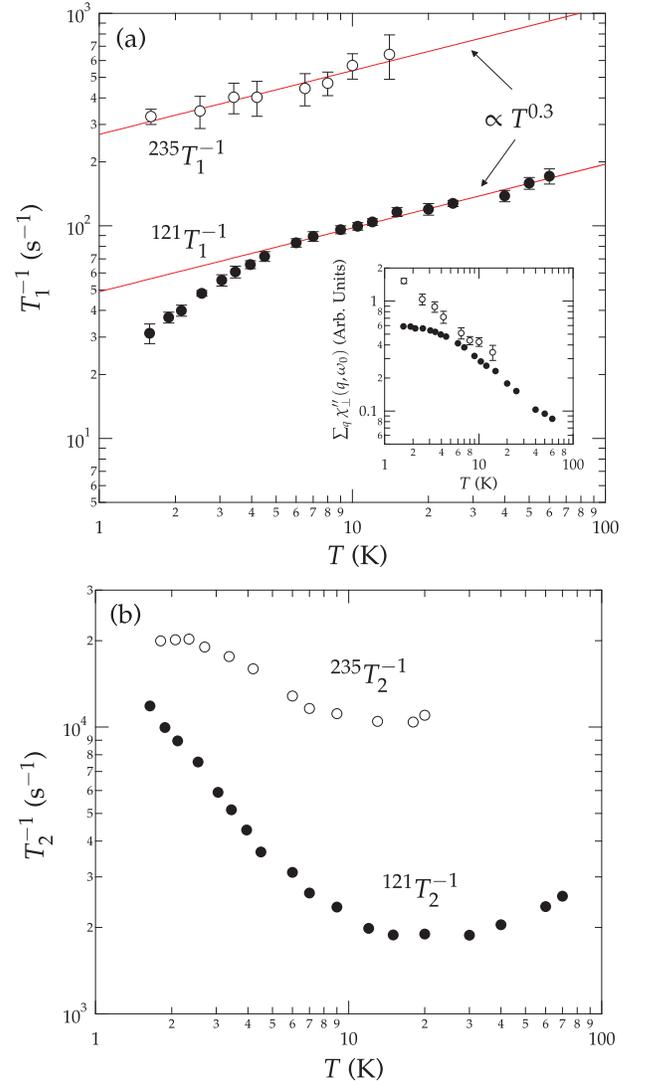}
\caption{Nuclear relaxation rates for both $^{235}$U and $^{121}$Sb(I).
(a) $T_1^{-1}$ as a function of $T$. Power law behaviors with the exponent 0.3 
were observed. However, $T_1^{-1}$ of $^{121}$Sb deviates from the power law 
below 5 K abruptly changing to $T_1^{-1}\propto T$. INSET: 
$\sum_q \chi_\perp''(q,\omega_0)$ vs.~$T$ which reflects the temperature 
variation of the spin fluctuations.  
(b) $T_2^{-1}$ for both nuclei are scaled according to $(\gamma_n A_\text{hf})^2$ 
exactly as $T_1^{-1}$ in the high temperature region above 5 K.  Below 5 K, 
$^{121}T_2^{-1}$ increases rapidly with  
decreasing temperature but $^{235}T_2^{-1}$ approaches a constant value.} \label{fig:T1T2} 
\end{figure}

For temperatures larger than about 5 K, $T_1^{-1}$ of both 
nuclei decreases slowly with
decreasing temperature, revealing the $T^{0.3}$ power law behavior. 
Below 5 K, however, $^{121}T_1^{-1}$ changes abruptly and
varies linearly with temperature, yet $^{235}T_1^{-1}$ shows no 
change down to 1.5 K.  A similar deviation in the temperature 
dependence of $T_2^{-1}$ is observed at 5 K, as shown in Fig.~4 (b), namely 
$^{121}T_2^{-1}$ increases rapidly but $^{235}T_2^{-1}$ 
increases slightly and saturates, with decreasing $T$.  
Interestingly, for U, both nuclear relaxation rates change somewhat
with temperature, but for the Sb, $T_1^{-1}$ decreases and $T_2^{-1}$ 
increases below $\sim 5$ K resulting in the fast increase of the ratio 
$T_2^{-1}/T_1^{-1}$ with decreasing $T$. 

In the ordered antiferromagnetic state, $T_1^{-1}$ is usually dominated 
by the fluctuations of the magnetic structure (magnons), in which a two magnon 
Raman process yields $T^3$ behavior.\cite{jaccarino66a}   Thus, the very small 
exponent of 0.3 in our case suggests that the relaxation mechanism in USb$_2$ is not 
governed by the simple magnon process.  
Although the origin of $T^{0.3}$ behavior 
is not clear, it suggests that the same 
relaxation mechanism is applicable to both on-site and ligand nuclear sites.  

The clear anomaly of both $^{121}T_1^{-1}$ and $^{121}T_2^{-1}$ at 5 K in contrast 
to those of $^{235}$U implies the dramatic change of the hyperfine coupling 
mechanism for $^{121}$Sb.
Since $A_\text{hf}$ for $^{235}$U is expected to be isotropic due to the 
overwhelming on-site Fermi contact term which is isotropic, the anisotropy of 
the spin fluctuations above $\sim 5$ K may be estimated using Eqs.~(\ref{eq:T1}) and 
(\ref{eq:T2}), i.e., $\sum_q\chi_\parallel''/\sum_q\chi_\perp''\sim 18$. When this 
ratio is applied for $^{121}$Sb, we obtain 
$A_\text{hf}^\parallel/A_\text{hf}^\perp\sim 1$. Therefore,  $A_\text{hf}$ is 
also apparently isotropic for $^{121}$Sb above 5 K.  
Since the anisotropy of the spin fluctuations does not change much for the 
$^{235}$U, the anomaly of the relaxation rates of the $^{121}$Sb indicates 
that the anisotropy of $A_\text{hf}$ is developed below 5 K and the ratio 
$A_\text{hf}^\parallel/A_\text{hf}^\perp$ increases with 
decreasing temperature up to 4 at 1.5 K. The anomalous behavior 
below 5 K may suggest that the hyperfine 
coupling on the Sb is very sensitive to even a small change of the electronic 
environment.  The otherwise slight increase of 
$^{235}T_2^{-1}$ below 5 K is then attributed to the cross-relaxation between $^{235}$U 
and $^{121}$Sb.

\section{Summary and conclusion}

The nuclear relaxation rates $T_1^{-1}$ and $T_2^{-1}$ of $^{235}$U are reported in the 
itinerant $5f$ electron system USb$_2$. The strong heating effect associated with the 
tiny gyromagnetic ratio of $^{235}$U prevents the direct measurement of 
$^{235}T_1^{-1}$, but we successfully accounted for the heating effect using two 
pulse sequences. The resultant $^{235}T_1^{-1}$ data as a function of temperature 
correctly scale according to $(\gamma_n A_\text{hf})^2$ with those of $^{121}$Sb, 
and vary as $T^{0.3}$. We find that $^{121}T_1^{-1}$ and $^{121}T_2^{-1}$ 
change dramatically at $\sim 5$ K, while $^{235}T_1^{-1}$ shows no change in the 
temperature dependence with the slight increase of $^{235}T_2^{-1}$.  
The different behavior is attributed to the different hyperfine coupling 
mechanism, but the origin of the anomaly at $\sim 5$ K is not clear at present. 
Nevertheless, the successful direct measurement of $T_1^{-1}$ on the $^{235}$U in an 
itinerant compound will pave the way for further direct investigations of the 
$^{235}$U nuclei in other U-based compounds.

\section*{Acknowledgment}

We thank the useful and delightful discussions with S. Kambe and H. Kato.  
This work was performed at Los Alamos National
Laboratory under the auspices of the US Department of
Energy Office of Science and suppored in part by the Laboratory Directed 
Research and Development program.

\bibliography{mybib}

\end{document}